# Metaheuristic Algorithm for Constrained Optimization in Radiation Therapy Treatment Planning: Design and Performance Comparison

Keshav Kumar K.[1*], Dr. NVSL Narasimham[2], Dr. A Rama Krishna Prasad[3]

[1] Assistant Professor, Department of Humanities and Mathematics, G. Narayanamma Institute of Technology and Science (for Women), Hyderabad-500 104, Telangana State, India.
Orcid ID: https://orcid.org/0000-0002-9211-2960

[2] Associate Professor, Department of Humanities and Mathematics, G. Narayanamma Institute of Technology and Science (for Women), Hyderabad-500 104, Telangana State, India.

[3] Retd.Professor, Department of Mathematics, Jawaharlal Nehru Technological University, Hyderabad-500 085, Telangana State, India.

Corresponding Author: keshav.gnits@gmail.com

*Abstract—* Radiation Therapy (RT) plays a pivotal role in the treatment of cancer, offering the potential to effectively target and eliminate tumor cells while minimizing harm to surrounding healthy tissues. However, the success of RT heavily depends on meticulous treatment planning that ensures the optimal balance between delivering a sufficiently high dose to the tumor and sparing nearby critical organs. This critical process demands a multidisciplinary approach that combines medical expertise, advanced imaging techniques, and computational tools. Optimization techniques have emerged as indispensable tools in refining RT planning, enabling the precise adjustment of radiation beam arrangements and intensities to achieve treatment objectives while adhering to strict dose constraints. This study delves into the realm of constrained optimization within RT Treatment Planning, employing metaheuristic algorithms to enhance the efficacy of this process. The research focuses on the design and performance comparison of three prominent optimization techniques: Bat Search Optimization (BSO), Bacterial Foraging Algorithm (BFA), and Artificial Bee Colony (ABC). Through systematic evaluation, it is observed that BFA exhibits superior execution time and convergence capabilities in comparison to the other algorithms. This research underscores the crucial importance of RT planning and highlights the imperative need for optimization methodologies to achieve optimal treatment outcomes.

## I. Introduction

Intensity Modulated Radiation Therapy (IMRT) stands as a widely adopted technique in the realm of RT [1]. The basis of radiotherapy involves directing localized radiation toward malignant tumor cells, causing damage to their DNA [2]. IMRT represents a sophisticated treatment approach, employing advanced external megavoltage radiation beams from linear accelerators with finely controlled modulations. Particularly effective for cancers like prostate and nasopharyngeal cancer, IMRT enables a concentrated and potent treatment of tumor regions, while keeping radiation exposure to nearby healthy tissues minimal [3]. This diverges from the traditional 3D conformal radiotherapy (3DCRT), which employs uniform large beams for treatment. During IMRT, the radiation dose varies across the treatment zone. To achieve this, the radiation source is broken up into pencil-sized beams, also called beamlets [4, 5]. In the process of planning RT, specific dose constraints are established for both the intended treatment area and the adjacent healthy structures. These constraints are typically not expressed as specific doses to individual points in the body, but rather as overall measures like maximum or minimum doses, and dose limits for specific volumes (dose-volume constraints). Subsequently, an inverse optimization process is carried out using numerical methods to determine the optimal intensities for each beamlet, based on the prescribed dose goals [6].

IMRT's primary benefit is that it allows for the Planning Target Volume (PTV) to receive the prescribed radiation dose in full, while healthy tissue and Organs at Risk (OARs) are protected to a greater extent [7]. To accomplish this, the radiation beams are formed to closely mimic the contours of the tumor, and the beams' spatial intensities are fine-tuned to account for the shape of the tumor and the potential danger to surrounding organs. A Multi-Leaf Collimator (MLC) is used to make this procedure easier. In IMRT, optimizing weight factors is central to the inverse planning idea. Beamlets are tiny, rectangular segments of the radiation field that are used in beamlet-based inverse planning (BBIP). These beamlet intensities, also called fluences, are the optimization variables, and they can number in the thousands. By mathematically formulating the IMRT optimization problem, we can guarantee that the tumor target always receives the correct dose while still respecting the established dose limitations for OARs in a clinical environment.

For the purpose of developing effective RT treatment regimens, two broad classes of optimization algorithms have been developed. In the first group, we find deterministic algorithms, which always proceed in the same way from the same initial state. They struggle to break out of local minimum points, however. The Newton-Raphson and the Nelder-Mead method are two examples of gradient-based and non-gradient-based algorithms, respectively. IMRT and volumetric arc therapy [8] have both benefited from these methods. The second type is heuristic algorithms, which adopt a different, unpredictable course with each iteration of the optimization process. Because of this trait, they can avoid getting stuck in a rut by employing exploration and exploitation strategies [9].

This research focuses on enhancing RT Treatment Planning through the utilization of three distinct metaheuristic optimization techniques. The paper is structured as follows: Section I introduces the study's objectives, while Section II conducts a comprehensive literature survey to establish the research's context. Section III details the methodology, outlining how the optimization techniques were employed in the context of RT Treatment Planning. In Section IV, the outcome of the optimization techniques are presented and discussed. The study concludes in Section V, summarizing key findings and potential future research directions.

II. LITERATURE SURVEY

This research draws from a thorough examination of recent academic papers that have informed its development. Several of these pivotal papers are outlined below, contributing significantly to the study's foundation and direction.

In the journal [10], the authors elaborate on the creation, execution, and resolution of a dependable direct aperture optimization model intended for IMRT planning in the context of cancer treatment. Their noteworthy contribution to the realm of operations research is manifest in the introduction of an innovative mixed-integer programming model that integrates both mechanical and clinical prerequisites pertinent to contemporary treatment equipment. Due to the model's complexity, they propose a novel heuristic to generate feasible treatment plans efficiently, demonstrated on five clinical patient datasets. In the study [11], an automated approach to treatment planning is presented, introducing the MetaPlanner (MP) algorithm. This algorithm automates the planning process by optimizing treatment planning hyperparameters through meta-optimization. Utilizing a derivative-free method, the MP algorithm searches for weight configurations that minimize a meta-scoring function. This function replicates clinical decision-making by prioritizing factors such as dose uniformity, conformity, spillage, and sparing of OARs. The algorithm's effectiveness is assessed using clinical datasets for both IMRT and Volumetric Modulated Arc Therapy (VMAT) planning. The results demonstrate comparable or superior outcomes compared to manually conducted VMAT planning. The idea of superiorization in IMRT treatment planning is introduced in the journal [12]. To direct a feasibility-seeking projection method toward a feasible point, superiorization uses linear voxel dose inequality constraints. A nonlinear objective function is optimized using gradient descent steps. The matRad toolbox is used to apply this strategy, which provides a basis for improvement. This framework's effectiveness in IMRT treatment planning is evaluated in comparison to that of feasibility-seeking and nonlinearly constrained optimization. Direct Aperture Optimization (DAO) in IMRT planning is given a mixed-integer nonlinear mathematical framework in the journal [13]. Metaheuristic techniques like Differential Evolution (DE) and Particle Swarm Optimization (PSO) are used to take on this problem's complexity. Taguchi is used to find optimal values for the algorithm parameters. PSO surpasses DE by means of treatment quality and computing efficiency, according to tests conducted on real patient data involving liver tumor.

Paper [14] introduces a novel approach called direct angle and aperture optimization, which integrates decisions related to beam directions, intensities, and aperture shapes. Metaheuristic algorithms based on DE are developed to solve this problem, and their parameters are optimized using the Taguchi design of experiments. Evaluations conducted on liver cancer cases demonstrate the superiority of one specific algorithm, ahdDE-PSO, in generating high-quality treatment plans. The researcher [15] attempts to solve the problem of IMRT planning's lack of diversity in solution generation by posing it as a massive combinatorial many-objective optimization issue. A co-evolutionary technique is developed, which combines fine and rough encoding for local exploitation and global exploration. Customized local search algorithms are implemented. The suggested approach outperforms state-of-the-art algorithms and converges faster, according to the experimental findings. In the study [16] we look into how contemporary Bayesian Optimization (BO) techniques can be used in robotic therapy planning. In a sample of IMRT treatment cases, BO techniques are evaluated and compared to clinical plans and other optimization strategies for both quality and efficiency. This research compares several different BO techniques, including Gaussian Process with Expected Improvement (GPEI) and Sparse Axis Aligned Subspace BO (SAAS-BO). An Intelligent Treatment Planner Network (ITPN) is presented in the article [17] that uses Deep Reinforcement Learning for treatment planning. The ITPN modifies dwell durations in a way that mimics how humans make decisions in order to improve the quality of plans. The network is trained to optimize dwell periods in accordance with a hybrid equivalent uniform dosage objective function. ITPN's superior performance over IPSA is demonstrated by comparing the two methods side by side.

III. METHODS

This section provides an extensive overview of the methodology employed in this investigation. It covers a wide range of topics, including Treatment Planning Preparation, constraint handling, and the utilization of optimization techniques.

*A. Treatment Planning Preparation*

In our study, we employed the Computational Environment for Radiotherapy Research (CERR) [18] as the treatment planning system. CERR is a software built on MATLAB® that provides a comprehensive package for RT treatment planning. Merits of CERR include its user interface, access to matrices describing dosage deposition, the integration of specialized programming modules, and the availability of MATLAB toolboxes. Based on the specified target and structures of the anatomy, the planner chooses the number and orientation of 6 MV photon beams for treatment with the help of the planning CT scan.

Beams are deliberately arranged in an environment of conformal therapy with uniform radiation fields to avoid beam collisions and reduce overlap with normal OARs. Beams in IMRT plans are broken down into a series of smaller units called "beamlets," normally measuring 0.5 cm × 0.5 cm but can be as small as 0.2 cm × 0.2 cm. The intensity of these beamlets is then determined using an optimization method. The primary objective of this optimization approach is to maximize the dose delivered to the PTV while restricting the dose absorbed by OARs. Figure 1 provides an illustration of a treatment approach for prostate cancer, utilizing a single anterior beam. The beam's upper section illustrates the intensity distribution of its constituent beamlets.

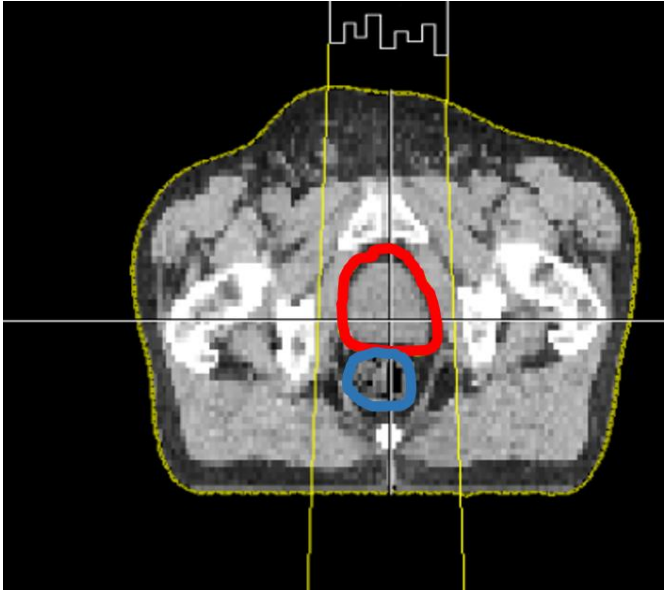

Fig. 1. IMRT plane for the prostate as seen on an axial CT scan.

During the process of IMRT treatment planning, it's necessary to define a two-dimensional photon fluence map for each beam. This fluence map consists of controllable individual beamlet intensities. Subsequently, the optimization process involves calculating the dose delivered to each voxel within all structures, thereby providing optimization feedback.

In the realm of IMRT optimization, the specifications are formulated as follows: a prescription dose for the PTV is required, while adhering to maximum dose restrictions for OARs.

$$f(x) = \arg\min \sum_{i \in T} \frac{(D_i - D_{PTV})^2}{N_T}, \qquad i \in PTV \qquad [1]$$

$$s.t \ maxD_j \leq D_{max,OAR_j}, \qquad \forall j \in OAR_j \qquad [2]$$

Here, let $N_T$ represent the total voxels within the target structure, also referred to as the $PTV$. $D_{PTV}$ signifies the recommended dose assigned to the target structure. Within this context, $D_j$ denotes the dose of each specific voxel labelled as "j," while $D_{max,j}$ stands for the maximum permissible dose designated for the respective $OAR_j$. Additionally, "$D_i$" corresponds to the dose received by voxel "$i$" within the PTV. Equation 1 captures the extent to which the PTV is covered by the required prescription dose, $D_{PTV}$. The dose attributed to voxel "$i$" demonstrates a linear relationship with the Dose Deposition Matrix (DDM):

$$D_i = \sum_{k=1}^{M} \sum_{j=1}^{N} A_{k,j}^i \cdot x_{k,j}^i \qquad [3]$$

Within this framework, $A_{k,j}$ represents the DDM that characterizes the contribution of dose to pertinent voxels within the structure of interest for unit fluence. The index "$k$" signifies the beam number, while $x_{k,j}$ represents the beamlet intensity "$j$" for beam "$k$". Notably, "$M$" and "$N$" denotes the total count of beams, and beamlets associated with "$k$". At its core, the process of IMRT treatment planning can be seen as an optimization problem aimed at minimization. The optimization process revolves around the continuous fluence intensities represented as $x_{k,j}$ for the beamlets. These fluence intensities serve as the decision variables in this context. The DDM was computed using the ORART (Operations Research Applied to RT) toolbox, a CERR extension. The computation relies on Ahnesjo's pencil beam approach [19].

*B. Optimization Algorithms*

Here, we introduce the utilization of metaheuristic algorithms to tackle the challenges of constrained optimization in the realm of RT treatment planning. This approach is aimed at elevating the effectiveness of IMRT planning by optimizing the arrangement of radiation beams. The goal is to achieve precise tumor targeting while concurrently minimizing the impact on adjacent healthy tissues. The working of optimization techniques are detailed below.

**Bat Search Optimization (BSO):** BSO, inspired by the behaviours of bats in their search for prey [20], is employed to tackle the complexities of constrained optimization in this medical context. The algorithm mimics the echolocation and movement patterns of bats to navigate through the search space and converge toward optimal solutions. The procedure of BSO for constrained optimization in RT treatment planning can be outlined as follows:
- The optimization process begins with the initialization of a population of virtual "bats" [21]. Each bat represents a

potential solution in the search space, which corresponds to different combinations of beamlet intensities.
- Just as bats emit ultrasonic signals for navigation, each bat in the algorithm sends out pulses that represent its potential solution. These pulses are adjusted according to a specified loudness and pulse rate. Bats with louder pulses are more likely to find better solutions.
- Bats perform both local and global searches [22]. In the local search, bats explore around their current positions to identify promising areas. In the global search, some bats randomly explore the entire search space to enhance diversity and discover potentially better solutions.
- After exploring, bats adjust their positions towards more promising solutions. This adjustment is influenced by the best solutions found so far, as well as the echolocation process.
- One of the key challenges in RT treatment planning is dealing with constraints. BSO handles constraints by penalizing infeasible solutions, ensuring that the optimization process adheres to the specified dose constraints and other physical limitations.
- The above steps are iteratively executed to allow bats to refine their solutions over multiple generations. The optimization process terminates after a predefined number of iterations or when convergence criteria are met.

**Artificial Bee Colony Optimization (ABC):** ABC is inspired by the foraging behaviour of honeybee colonies [23]. It stimulates the process of honeybee scouts searching for food sources and sharing information with other bees in the hive. The algorithm leverages this natural behaviour to navigate the solution space and identify optimal solutions that fulfil both dose requirements and physical constraints. The working process of ABC Optimization for constrained optimization in RT treatment planning can be explained below:
- The optimization starts with an initial population of artificial bees, each representing a potential solution. These solutions correspond to various configurations of beamlet intensities [24].
- In this phase, employed bees evaluate the quality of their solutions based on an objective function, which integrates both the prescribed dose to the target and constraints related to maximum doses at OARs. Bees adjust their solutions by exploring nearby regions in the solution space.
- Other bees, referred to as onlooker bees, select their solutions based on the quality information shared by employed bees. Better-performing solutions are more likely to be chosen, while solutions that do not adhere to constraints are less favoured.
- Occasionally, a solution may not yield improvement for an extended period. In such cases, the solution is abandoned, and the corresponding bee becomes a scout bee. Scout bees then explore new, unexplored areas of the solution space.
- ABC Optimization addresses constraints by adjusting the selection probabilities of solutions based on their feasibility [25]. Infeasible solutions are penalized to encourage adherence to dose limits and other physical constraints.
- The optimization process iteratively progresses through the employed bees, onlooker bees, and scout bees phases. This iterative approach allows the algorithm to refine solutions over multiple cycles. The optimization process reaches its conclusion either after a specified number of iterations have been completed or when the predefined convergence criteria have been satisfied.

**Bacterial Foraging Algorithm (BFA):** BFA draws inspiration from the foraging behaviour of bacteria in search of nutrients [26]. It models the interactions between bacteria and their environment to navigate through solution spaces and locate optimal solutions that satisfy both dose requirements and physical constraints. The operational process of Bacterial Foraging Optimization for constrained optimization in RT treatment planning can be outlined as follows:
- The algorithm begins with an initial population of virtual bacteria, each representing a potential solution [27]. These solutions represent different arrangements of beamlet intensities.
- Bacteria undergo chemotactic movement, which simulates their exploration of the environment for nutrients. Solutions are adjusted based on the concept of bacterial movement toward areas with higher nutrient concentrations. In the context of RT, nutrient concentration correlates with optimal solution quality.
- Bacteria that have explored regions with higher nutrient concentrations are more likely to reproduce, generating new solutions [28]. Meanwhile, bacteria that have not found promising solutions undergo elimination-dispersal, simulating the concept of bacteria dispersing when resources are scarce.
- Bacteria communicate and share information about their exploration. This fosters cooperation and enables the algorithm to collectively converge towards optimal solutions.
- BFA Optimization addresses constraints by incorporating them into the optimization process. Solutions that violate constraints are penalized, and the algorithm guides the exploration towards feasible regions of the solution space.
- The optimization process iterates through chemotactic movement, reproduction, elimination-dispersal, and communication phases. This iterative approach refines solutions over multiple generations. The optimization process concludes after a predetermined number of iterations or when convergence criteria are met.

*C. Constraints*

Constraints in the optimization problem are commonly divided into two categories: "geometrical" constraints and "physical" constraints. Limits on the range of the design variables might be thought of as upper and lower bounds for the geometric constraints. Contrarily, physical constraints impose bounds on physical quantities that can only be evaluated numerically. Equation 1 is a constrained optimization problem; an approach was used to prevent constraint violations by transforming the problem into an unconstrained one. This was accomplished by punishing computationally expensive infeasible objective function solutions [29] rather than attempting to correct them. Dose-based constraints for OARs were preferred in this investigation. Dose-based constraints were chosen because of their conceptual simplicity and widespread application in clinical practice [30]. The optimization problem with dose constraints used the following pseudo-objective function:

$$\emptyset(x) = f(x) + \sum_{i=1}^{N}\left[w_j H\left(maxD_{OAR_j} - D_{max,OAR_j}\right) \cdot maxD_{OAR_j}\right] \quad [4]$$

Here, "$x$" denotes the intensities of the beamlets, while "$f(x)$" and "$H(.)$" represents the quadratic dose and Heaviside function associated with the PTV. The index "$j$" encompasses the various OARs. "$maxD_{OAR_j}$" signifies the maximum permissible dose for the $OAR_j$, and "$D_{max,OAR_j}$" indicates the maximum dose at $OAR_j$, iteratively computed throughout the optimization phase. Additionally, "$w_j$" is a weight factor linked to the OARs, and "$N$" represents the total count of OARs. Then, a repair function "$g(x_i)$" was employed to ensure that each individual beamlet's intensity "$x_i$" falls within an acceptable range. This repair function addresses geometrical constraints, ensuring that the beamlet intensities remain within predefined limits. Specifically:

$$g(x_i) = \begin{cases} lower_{bound} & if \ x_i < lower_{bound} \\ upper_{bound} & if \ x_i > upper_{bound} \end{cases} \quad [5]$$

In this context, the lower and upper bounds were established through empirical experimentation, set at 0 and 20 respectively.

## IV. RESULT AND DISCUSSION

In this section, we evaluated the performance of three optimization techniques in the context of constrained optimization for RT treatment planning. The evaluation criteria included execution time and convergence properties.

### A. Execution Time

Figure 2 presents the absolute execution times of the metaheuristic algorithms in relation to the number of pencil beams (PBs) utilized. Among the three techniques – BFA, ABC, and BSA. BFA exhibited superior performance. Notably, the execution time of the optimization approaches demonstrated a linear relationship with the number of PBs.

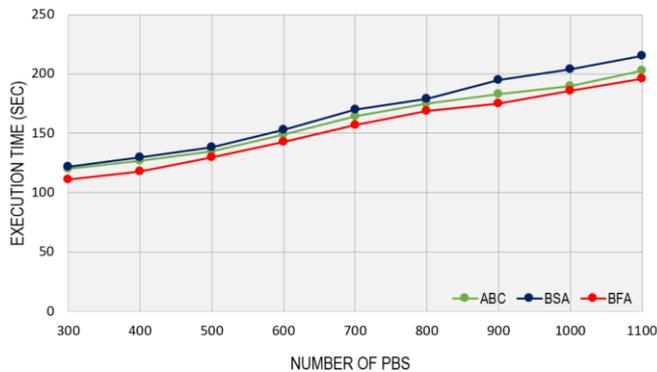

Fig. 2. Execution time comparison

### B. Convergence

Achieving convergence within the optimization process is critical for obtaining optimal solutions. The algorithms employed in this study – BSA, ABC, and BFA – demonstrated convergence to both local and global optima. The iterative nature of these algorithms facilitated their comparison against optimal solutions by assessing the best solutions at each iteration. In practical terms, the convergence of BSA, ABC, and BFA was consistently achieved within a modest number of iterations, generally fewer than 60. This indicates the effectiveness of these algorithms in navigating the solution space and identifying optimal solutions in a relatively short span of iterations. Figure 3 showcases the evolution of the objective function $\varphi(x)$ over 100 iterations for the three optimization techniques, focusing on a prostate case. It is evident that the BFA displayed a quicker convergence rate compared to the other methods. The blue plot corresponds to BSA's convergence, the red plot depicts BFA's convergence, and the green plot represents ABC's convergence.

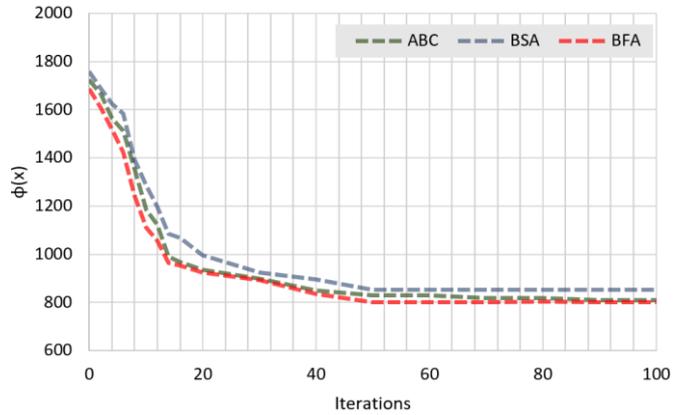

Fig. 3. Convergence comparison

### C. Correlation Between Weight Factors and Plan Quality

Based on the comparison of execution times and convergence properties, the BFA emerged as the optimal solution for treatment planning. Using the BFA optimization approach, we investigate the impact of penalizing weight factors ($w_j$) through the total quality of prostate cancer treatment plans. To perform a qualitative assessment of treatment plan quality, we adopted the cumulative Dose Volume Histogram (DVH), a commonly employed metric in the field of RT. The DVH serves to condense the three-dimensional dose distribution into a concise two-dimensional visual format.

Figure 4 displays the overall DVHs for the PTV when various penalizing factors, $w_B$ and $w_R$, on the bladder and rectum are used. The horizontal and the vertical axis indicates the desired dose level, and the volume of the PTV acquiring that dose or above. An ideal DVH for the PTV exhibits a steep drop-off, indicating excellent tumor coverage. The appearance of a step-like drop at the prescribed dose ($D_0$) signifies that the entire volume received the prescribed dose. For the PTV, this steep drop-off is crucial to ensure comprehensive tumor coverage. Moreover, dose constraints impact the DVH curve's shape for each Volume of Interest (VOI). It is important to note that optimizing a treatment plan involves balancing the quality of the DVHs for various VOIs. The optimizer may slightly compromise the DVH quality for some VOIs to enhance those of others. The $w_j$ in the optimization algorithm (Equation 4) indicate the amount of priority assigned for each dose constraint.

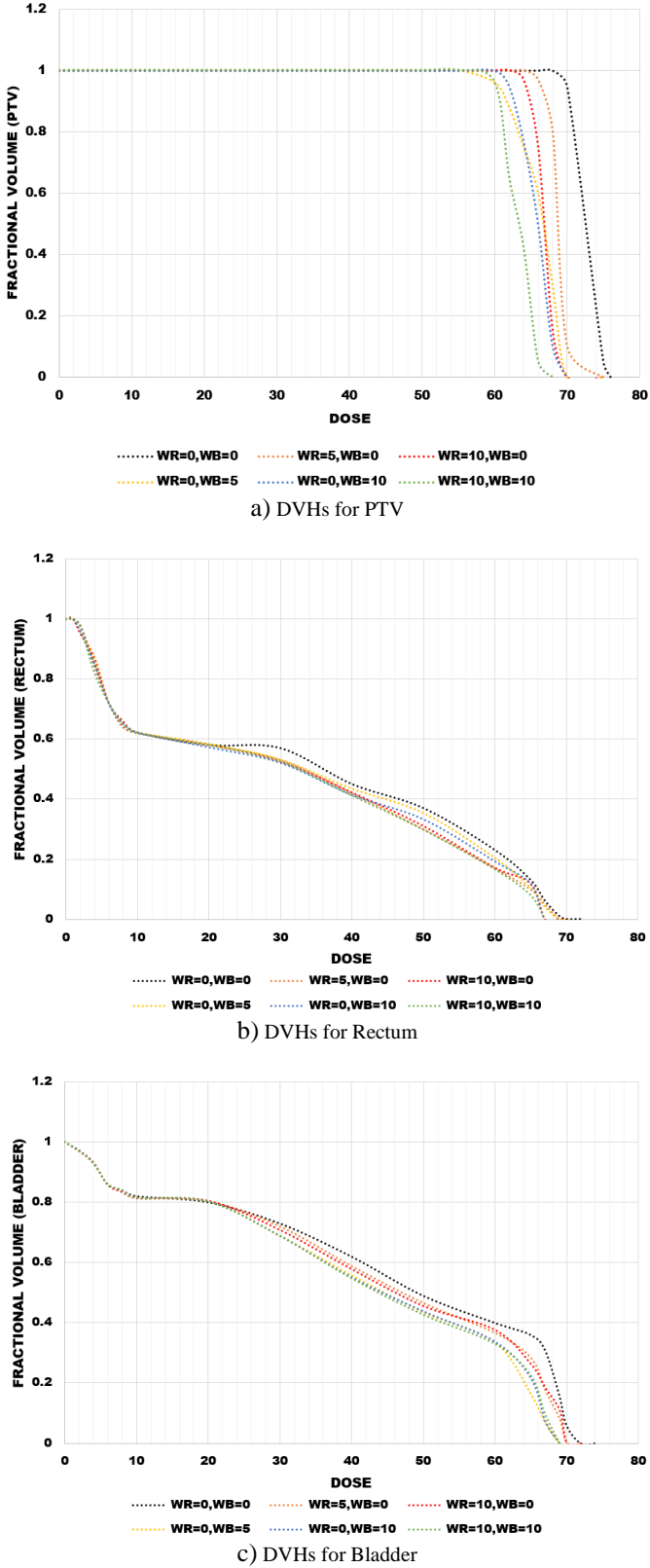

Fig. 4. DVHs with various combinations of penalizing weight factors.

Figure 4 (a) illustrates the DVH for the unconstrained plan ($w_R$=0, $w_B$=0), showing a sharp curve that signifies thorough coverage of the prostate cancer. However, as the $w_j$ are elevated, as demonstrated in Figures 4 (b) and (c), the PTV coverage becomes affected. Meanwhile, there is a reduction in the dose administered to OARs. This trend can be attributed to the intersection of the bladder and rectum volumes with the volume of tumor. In conclusion, this section delves into the significance of $w_j$ in influencing treatment plan quality. The utilization of cumulative DVHs sheds light on the impact of these factors on the trade-offs between optimal tumor coverage and adherence to dose constraints for various structures. This comprehensive analysis contributes to a deeper understanding of the interplay between penalizing factors and the resulting treatment plan characteristics.

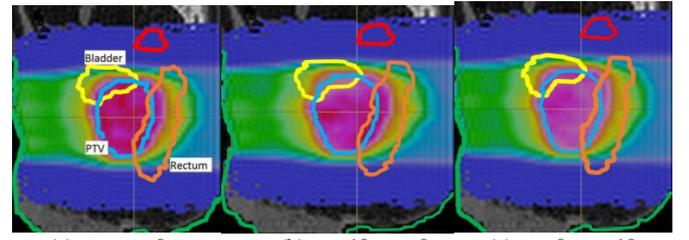

(a) $w_R$=$w_B$=0   (b) $w_R$=10, $w_B$=0   (c) $w_R$=0, $w_B$=10

Fig 5. Sagittal dose distribution view of unconstrained IMRT optimization.

Figure 5 provides a graphical representation of this phenomenon by displaying the dosage distribution in a sagittal plane for three distinct sets of weight components. It's worth highlighting that as the weight factor increases, the dose administered to the OARs decreases. Additionally, the PTV overlaps with both the bladder and rectum, as demonstrated in the image. The results of the experiments related to prostate cancer are detailed in Table 1. Adhering to the recommendations of ICRU-50 [31], it's important to emphasize that the presented findings have not been normalized to achieve the prescribed dose for the specified volume or point. Normalization involves a straightforward procedure in which the PTV dosage, along with other components, is progressively escalated for each set of penalized weights. The dose values are expressed in units of Gy.

Table 1. Performance metrics of PTV, Rectum, and Bladder.

| $(W_R, W_B)$ | $D_{95}^{PTV}$ | $D_{max}^{Rectum}$ | $D_{max}^{Bladder}$ |
|---|---|---|---|
| (0,0) | 65.7 | 72.8 | 71.6 |
| (5,0) | 63.5 | 67.3 | 68.9 |
| (10,0) | 61.2 | 67.1 | 68.6 |
| (0,5) | 59.8 | 70.8 | 61.3 |
| (0,10) | 59.6 | 70.5 | 60.9 |
| (10,10) | 59.5 | 67.2 | 60.7 |

## V. CONCLUSION

In conclusion, this study has showcased the significant potential of metaheuristic algorithms in addressing constrained optimization challenges within RT Treatment Planning. Through a comparative analysis of three optimization techniques, BFA emerged as a standout performer, characterized by superior execution time and convergence capabilities. Furthermore, the influence of penalizing weight factors on treatment plan quality was effectively illustrated using BFA. While higher weight factors enhance the sparing of OAR), they entail a trade-off with compromised PTV coverage. This intricate balance highlights the complex nature of RT treatment planning, where optimal outcomes must be achieved within safety constraints.

As a future avenue of exploration, further investigation could delve into hybrid approaches that amalgamate the strengths of different metaheuristic algorithms. Integrating the distinct advantages of various techniques could potentially yield even more optimized and efficient solutions for RT treatment planning. Additionally, the study could extend to accommodate more complex treatment scenarios, such as adaptive therapy planning, which involves dynamic adjustments to treatment plans based on real-time patient data. This direction holds promise for advancing the field and enhancing the efficacy of RT for the benefit of patients worldwide.